\documentclass[twocolumn,showpacs,preprintnumbers,amsmath,amssymb]{revtex4}
\usepackage{graphicx}
\usepackage{dcolumn}
\usepackage{bm}

 \newcommand{\refeq}[1]{~(\ref{#1})}
 \newcommand{\myref}[1]{~\ref{#1}}
 \newcommand{\Sl}{S--$\ell$}
 \newcommand{\EM}{e.m.}

\newcommand{\be}{\begin{equation}}
\newcommand{\ee}{\end{equation}}
\newcommand{\bea}{\begin{eqnarray}}
\newcommand{\eea}{\end{eqnarray}}

\newcommand{\hyperg}{\,\,\,\mathrm{F}\!\!\!\!\!\!_2\,\,\,_1}

\begin{document}

\title{L\'evy--Student Distributions for Halos in Accelerator Beams}

\author{Nicola Cufaro Petroni}
 \email{cufaro@ba.infn.it}
 \affiliation{Dipartimento di Matematica
dell'Universit\`a di Bari, and INFN Sezione di Bari, \\ via E.
Orabona 4, 70125 Bari, Italy}

\author{Salvatore De Martino}
 \email{demartino@sa.infn.it}
\author{Silvio De Siena}
 \email{desiena@sa.infn.it}
\author{Fabrizio Illuminati}
 \email{illuminati@sa.infn.it}
\affiliation{Dipartimento di Fisica dell'Universit\`a di Salerno,\\
INFM Unit\`a di Salerno and INFN Sezione di Napoli -- Gruppo
collegato di Salerno,
\\ Via S. Allende, I--84081 Baronissi (SA), Italy}

\date{October 28, 2005}

\begin{abstract}
We describe the  transverse beam distribution in particle
accelerators within the controlled, stochastic dynamical scheme of
the Stochastic Mechanics (SM) which produces time reversal invariant
diffusion processes. This leads to a linearized theory summarized in
a Shchr\"odinger--like (\Sl)\ equation. The space charge effects
have been introduced in a recent paper~\cite{prstab} by coupling
this \Sl\ equation with the Maxwell equations. We analyze the space
charge effects to understand how the dynamics produces the actual
beam distributions, and in particular we show how the stationary,
self--consistent solutions are related to the (external, and
space--charge) potentials both when we suppose that the external
field is harmonic (\emph{constant focusing}), and when we \emph{a
priori} prescribe the shape of the stationary solution. We then
proceed to discuss a few new ideas~\cite{epac04} by introducing the
generalized Student distributions, namely non--Gaussian, L\'evy
\emph{infinitely divisible} (but not \emph{stable}) distributions.
We will discuss this idea from two different standpoints: (a) first
by supposing that the stationary distribution of our (Wiener
powered) SM model is a Student distribution; (b) by supposing that
our model is based on a (non--Gaussian) L\'evy process whose
increments are Student distributed. We show that in the case (a) the
longer tails of the power decay of the Student laws, and in the case
(b) the discontinuities of the L\'evy--Student process can well
account for the rare escape of particles from the beam core, and
hence for the formation of a halo in intense beams.
\end{abstract}


\maketitle

\section{Introduction}\label{introduction}

In high intensity beams of charged particles, proposed in recent
years for a wide variety of accelerator--related applications, it is
very important to keep at low level the beam loss to the wall of the
beam pipe, since even small fractional losses in a high--current
machine can cause exceedingly high levels of radioactivation. It is
now widely believed that one of the relevant mechanisms for these
losses is the formation of a low intensity beam halo more or less
far from the core. These halos have been observed \cite{koziol} or
studied in experiments \cite{reiser}, and have also been subjected
to an extensive simulation analysis \cite{gluck}. For the next
generation of high intensity machines it is however still necessary
to obtain a more quantitative understanding not only of the physics
of the halo, but also of the beam transverse distribution in general
\cite{hofmann}. In fact ``because there is not a consensus about its
definition, halo remains an imprecise term'' \cite{wangler} so that
several proposals have been put forward for its description.

The charged particle beams are usually described in terms of
classical, deterministic dynamical systems. The standard model is
that of a collisionless plasma where the corresponding dynamics is
embodied in a suitable phase space (see for example~\cite{landau}).
In this framework the beam is studied by means of the
\emph{particle--in--core} (\emph{pic}) model and the simulations
show that the instabilities due to a parametric resonance can allow
the particles to escape from the core with consequent halo
formation~\cite{gluck,hofmann,wangler}. The present paper takes a
different approach: it follows the idea that the particle
trajectories are samples of a stochastic process, rather than usual
deterministic (differentiable) trajectories. In the usual dynamical
models there is a particle probability distribution obeying the
Vlasov equation, and its evolution is Liouvillian in the sense that
the origin of the randomness is just in the initial conditions:
along the time evolution, which is supposed to be deterministic,
there is no new source of uncertainty. It is the non linear
character of the equations which produces the possible unpredictable
character of the trajectories. On the other hand in our model the
trajectories are replaced by stochastic processes since the time
evolution is supposed to be randomly perturbed even after the
initial time. It is open to discussion which one of these two
description is more realistic; in particular we should ask if the
mutual interactions among the beam particles look like random
collisions, or rather like continuous deterministic interactions. In
the opinion of the authors, however, a plasma (with collisions)
described in terms of controlled stochastic processes is a good
candidate to explain the rare escape of particles from a
quasi--stable beam core by statistically taking into account the
random inter--particle interactions that can not be described in
detail. Of course the idea of a stochastic approach is hardly
new~\cite{landau,ruggiero}, but there are several different ways to
implement it.

First of all let us remark that the system we want to describe is
endowed with some measure of invariance under time reversal, since
the external fields act to keep it in a quasi--stationary non
diffusive state despite the repulsive electro--magnetic (\EM)\
interactions among the constituent particles. However, a widespread
misconception notwithstanding, a theory of stochastic processes not
always describe irreversible systems: the addition of a dynamics to
a stochastic kinematics can in fact ascribe a measure of time
reversal invariance also to a stochastic system~\cite{paul}. The
standard way to build a stochastic dynamical system is to modify the
phase space dynamics by adding a Wiener noise $\mathbf{B}(t)$ to the
momentum equation only, so that the usual relations between position
and velocity is preserved:
\begin{eqnarray*}
m\,d{\bf Q}(t)&=&{\bf P}(t)\,dt\,,\\
d{\bf P}(t)&=&{\bf F}(t)\,dt+\beta\, d{\bf B}(t)\,.
\end{eqnarray*}
In this way we get a derivable, but not Markovian position process
${\bf Q}(t)$. The standard example of this approach is that of a
Brownian motion in a force field described by an Ornstein--Uhlenbeck
system of stochastic differential equations (SDE)~\cite{nelson}.
Alternatively we can add a Wiener noise ${\bf W}(t)$ with diffusion
coefficient $D$ to the position equation:
\[
d{\bf Q}(t)\;=\;{\bf v}_{(+)}({\bf Q}(t),t)\,dt+\sqrt{D}\,d{\bf
W}(t)\,.
\]
and get a Markovian, but not derivable ${\bf Q}(t)$. In this way the
stochastic system is also reduced to a single SDE since we are
obliged to drop the second (momentum) equation: in fact now ${\bf
Q}(t)$ is no more derivable. The standard example of this reduction
is the Smoluchowski approximation of the Ornstein--Uhlenbeck process
in the overdamped case~\cite{nelson}. As a consequence we will work
only in a configuration, and not in a phase space; but this does not
prevent us from introducing a dynamics -- as we will show in the
Section\myref{stochmech} -- either by generalizing the Newton
equations~\cite{nelson,guerraphysrep}, or by means of a stochastic
variational principle~\cite{guerravariaz}. Remark that in this
scheme the forward velocity ${\bf v}_{(+)}({\bf r},t)$ can no more
be an a priori given field: rather it now plays the role of a new
dynamical variable of our system. This second scheme, the Stochastic
mechanics (SM), is universally known for its original application to
the problem of building a classical stochastic model for Quantum
Mechanics (QM), but in fact it is a very general model which is
suitable for a large number of stochastic dynamical
systems~\cite{paul,albeverio}. We will also see in the
Section\myref{stochmech} that from the stochastic variational
principles two coupled equations are derived which are equivalent to
a Schr\"odinger--like (\Sl) differential equation: in this sense we
will speak of quantum--like (Q-$\ell$) systems, in analogy with
other recent researches on this subject~\cite{fedele,pusterla}. In
fact the SM can be used to describe every stochastic dynamical
system satisfying fairly general conditions: it is known since
longtime~\cite{morato}, for example, that for any given diffusion
there is a correspondence between diffusion processes and solutions
of \Sl\ equations where the Hamiltonians come in general from
suitable vector potentials. Under some regularity conditions this
correspondence is seen to be one-to-one. The usual Schr\"odinger
equation, and hence QM, is recovered when the diffusion coefficient
coincides with $\hbar/2m$, namely is connected to the Planck
constant. However we are interested here not in a stochastic model
of QM, but in the description of particle beams.

In the present paper we intend to widen the scope of our SM model by
introducing the idea that an important role for the beam dynamics
can be played by non--Gaussian L\'evy distributions. In fact these
distributions enjoyed a widespread popularity in the recent years
because of their multifaceted possible applications to a large set
of problems from the statistical mechanics to the mathematical
finance (see for example~\cite{paul,mantegna} and references quoted
therein). In particular the so called \emph{stable} laws (see
Section\myref{idstlaws}) are used in a large number of instances, as
for example in the definition of the so--called L\'evy flights. Our
research is instead focused on a family of non--Gaussian L\'evy laws
which are \emph{infinitely divisible} but not stable: the generalize
Student laws. As will be discussed later this will allow us to
overcome  -- without resorting to the trick of the truncated laws --
the problems raised by the fact that the stable non--Gaussian laws
always have divergent variances: a feature which is not realistic to
ascribe to most real systems. It is possible to show indeed that by
suitably choosing the parameters of the Student laws we can have
distributions with finite variance, and approximating the Gaussian
law as well as we want. On the other hand the infinitely divisible
character of these laws is all that is required to build a
stationary, stochastically continuous Markov process with
independent increments, namely the L\'evy process that we propose to
use to represent the evolution of our particle beam.

Of course it is not always mathematically easy to deal with the
infinitely divisible processes, but we will show that at least in
two respects they will help us to have some further insight in the
beam dynamics. First of all we use the Student distributions in the
framework of the traditional SM where the randomness of process is
supplied by a Gaussian Wiener noise: here we examine the features of
the self--consistent potentials which can produce a Student
distribution as stationary transverse distribution of a particle
beam. In this instance the focus of our research is on the increase
of the probability of finding the particles at a great distance from
the beam core. Then we pass to the definition of a true
\emph{L\'evy--Student process}, and we show with a few simulations
that these processes can help to explain how a particle can be
expelled from the bunch because of some kind of hard collision. In
fact the trajectories of our L\'evy--Student process show the
typical jumps of the non--Gaussian L\'evy processes: a feature that
we propose to use as a model for the halo formation. It is worth
remarking that, albeit the more recent empirical data about
halos~\cite{allen} are still not accurate enough to distinguish
between the suggested distributions and the usual Gaussian ones, our
conjecture on the role of Student laws in the transverse beam
dynamics has recently found a first confirmation~\cite{vivoli} in
numerical simulations showing how these laws are well suited to
describe the statistics of the random features of the particle
paths.

In a few previous papers~\cite{pre} we connected the (transverse)
r.m.s.\ emittance to the characteristic microscopic scale and to the
total number of the particles in a bunch, and implemented a few
techniques of active control for the dynamics of the beam. In this
paper we first of all review the theoretical basis~\cite{prstab,pre}
of the proposed model: in the Section~\ref{nelsonism} we define our
SM model with emphasis added on the potentials which control the
beam dynamics and on the possible non stationary solutions of this
model~\cite{ijmpb}. In the Section~\myref{selfconsistent} we review
our analysis of the self--consistent, space charge effects due to
the \EM\ interaction among the particles, adding a few new results
and comments. In the Section\myref{selfconspot} we then discuss the
idea~\cite{epac04} that the laws ruling the transverse distribution
of particle beams are non--Gaussian, infinitely divisible, L\'evy
laws as the generalized Student laws. In particular we analyze the
behavior of our usual SM model under the hypothesis that the
stationary transverse distribution is a Student law. Finally in the
Section\myref{studentprocess} we study the possibility of extending
our SM model to L\'evy processes whose increments are distributed
according to the Student law. We think in particular that the
presence of isolated jumps in the trajectories can help to build a
realistic model for the possible formation of halos in the particle
beams. We end the paper with a few conclusive remarks.

\section{Stochastic beam dynamics}\label{nelsonism}

\subsection{Stochastic mechanics}\label{stochmech}

First of all we introduce the stochastic process performed by a
representative particle that oscillates around the closed ideal
orbit in a particle accelerator. We consider the 3--dimensional
(3--DIM) diffusion process ${\bf Q} (t)$, taking the values
$\mathbf{r}$, which describes the position of the representative
particle and whose probability density is proportional to the
particle density of the bunch. As stated in the
Section\myref{introduction} the evolution of this process is ruled
by the It\^o stochastic differential equation (SDE)
\begin{equation}\label{ito}
 d{\bf Q}(t) = {\bf v}_{(+)}({\bf Q}(t), t)\,dt +
\sqrt{D}\,d{\bf W} (t) \, ,
\end{equation}
where ${\bf v}_{(+)}(\mathbf{r}, t)$ is the forward velocity, and
$d{\bf W} (t) \equiv {\bf W} (t + dt) - {\bf W} (t)$ is the
increment process of a standard Wiener noise $\mathbf{W}(t)$; as it
is well known this increment process is gaussian with law
$\mathcal{N}(0,\mathbb{I}\,dt)$, where $\mathbb{I}$ is the
$3\times3$ identity matrix. Finally the diffusion coefficient $D$ is
supposed to be constant: the quantity $\alpha=2mD$, which has the
dimensions of an action, will be later connected to the
characteristic transverse emittance of the beam. The
equation~(\ref{ito}) defines the random kinematics performed by the
particle, and replaces the usual deterministic kinematics
\begin{equation}\label{classicalkin}
 d{\bf q}(t) = {\bf v}(\mathbf{q}(t),t) dt
\end{equation}
where $\mathbf{q}(t)$ is just the trajectory in the 3--DIM space.

To counteract the dissipation due to this stochastic kinematics, a
dynamics must be independently added. In SM we do not have a phase
space: our description is entirely in a 3-DIM configuration space.
This means in particular that the dynamics is not introduced in a
Hamiltonian way, but by means of a suitable stochastic least action
principle~\cite{guerravariaz} obtained as a generalization of the
variational principle of classical mechanics. In the following we
will briefly review the main results, referring for details to the
references~\cite{paul,guerravariaz,nelson}. Given the
SDE~(\ref{ito}), we consider the probability density function (pdf)
$\rho({\bf r}, t)$ associated to the diffusion ${\bf Q}(t)$ so that,
besides the forward velocity ${\bf v}_{(+)}({\bf r}, t)$, we can now
define a backward velocity
\begin{equation}\label{vback}
 {\bf v}_{(-)}({\bf r}, t)  = {\bf
v}_{(+)}({\bf r}, t) - 2 D \frac{\nabla \rho ({\bf r}, t)}{\rho
({\bf r}, t)}\,.
\end{equation}
We can then introduce also the current and the osmotic velocity
fields, defined as:
\begin{equation}\label{osmotic}
{\bf v}=\frac{{\bf v}_{(+)} + {\bf v}_{(-)}}{2} \; ;\qquad {\bf
u}=\frac{{\bf v}_{(+)}-{\bf v}_{(-)}}{2}\, =\, D\frac{\nabla
\rho}{\rho} \, .
\end{equation}
Here ${\bf v}$ represents the velocity field of the density, while
$\bf u$ is of intrinsic stochastic nature and is a measure of the
non differentiability of the stochastic trajectories.

A first consequence of the stochastic generalization of the least
action principle~\cite{guerravariaz,nelson} is that the current
velocity takes the following irrotational form:
\begin{equation}\label{gradient}
m{\bf v} ({\bf r}, t) = \nabla S ({\bf r}, t) \, ,
\end{equation}
while the Lagrange equations of motion for the density $\rho$ and
for the current velocity ${\bf v}$ are the continuity equation
associated to every stochastic process
\begin{equation}\label{continuity}
\partial_{t} \rho = -\nabla \cdot (\rho {\bf v})\,,
\end{equation}
and a dynamical equation
\begin{equation}\label{hjm}
\partial_{t} S + \frac{m}{2} {\bf v}^{2} - 2m D^2
\frac{\nabla^{2} \sqrt{\rho}}{\sqrt{\rho}} + V({\bf r}, t) = 0 \,,
\end{equation}
which characterizes our particular class of time--reversal invariant
diffusions (Nelson processes). The last equation has the same form
of the Hamilton--Jacobi--Madelung (HJM) equation, originally
introduced in the hydrodynamic description of quantum mechanics by
Madelung~\cite{madelung}. Since~(\ref{gradient}) holds, the two
equations\refeq{continuity} and\refeq{hjm} can be put in the
following form
\begin{eqnarray}
 \partial_{t} \rho &=& -\frac{1}{m}\nabla \cdot (\rho \nabla S)\label{continuity1}\\
 \partial_{t} S &=&- \frac{1}{2m} {\nabla S}^{2} + 2m D^2
              \frac{\nabla^{2} \sqrt{\rho}}{\sqrt{\rho}} - V({\bf
                    r},t)\label{hjm2}
\end{eqnarray}
which now constitutes a coupled, non linear system of partial
differential equations for the pair $(\rho, S)$ which completely
determines the state of our beam. On the other hand, because
of\refeq{gradient}, this state is equivalently given by the
pair$(\rho, \mathbf{v})$.

It can also be shown by simple substitution from~(\ref{osmotic})
that~(\ref{continuity}) is equivalent to the standard Fokker--Planck
(FP) equation
\begin{equation}\label{fp}
\partial_{t} \rho = -\nabla \cdot [{\bf v}_{(+)} \rho]
+  D \, \nabla^2  \rho
\end{equation}
formally associated to the It\^o equation\refeq{ito}. In fact also
the HJM equation\refeq{hjm} can be cast in a form based on
$\mathbf{v}_{(+)}$ rather than on $\mathbf{v}$, namely
\begin{eqnarray}\label{hjm1}
 \partial_{t} S &=&-\frac{m}{2}\mathbf{v}_{(+)}^2+ m D\,\mathbf{v}_{(+)}\nabla\ln f
               \nonumber\\
 &&\qquad\qquad\qquad  +mD^2\nabla^2\ln f-V
\end{eqnarray}
where  $f$ is a dimensionless density defined by
\begin{equation}\label{dimensionlesspdf}
 \rho(\mathbf{r},t) = C f(\mathbf{r},t)
\end{equation}
where $C$ is a dimensional constant. On the other hand,
from\refeq{vback} and\refeq{osmotic}, we know that also the forward
velocity $\mathbf{v}_{(+)}$ is irrotational:
\begin{equation}\label{gradient1}
 \mathbf{v}_{(+)}(\mathbf{r},t)=\nabla W(\mathbf{r},t)\,,
\end{equation}
and that by taking\refeq{gradient} into account the functions $W$
and $S$ are connected by the relation
\begin{equation}\label{phases}
S(\mathbf{r},t)=mW(\mathbf{r},t)-mD\ln f(\mathbf{r},t) -\theta(t)
\end{equation}
where $\theta$ is an arbitrary function of $t$ only.

The time--reversal invariance is now made
possible~\cite{guerraphysrep} by the fact that the forward drift
velocity ${\bf v}_{(+)} ({\bf r}, t)$ is no more an {\it a priori}
given field, as is usual for the diffusion processes of the Langevin
type; instead it is dynamically determined at any instant of time,
starting by an initial condition, through the HJM evolution
equation~(\ref{hjm}). It is finally important to remark that,
introducing the representation~\cite{madelung}
\begin{equation}\label{schroedpsi}
\Psi ({\bf r}, t) = \sqrt{\rho({\bf r}, t)}\, {\mathrm e}^{i S({\bf
r}, t)/\alpha} \, ,
\end{equation}
(with $\alpha=2mD$) the coupled equations~(\ref{continuity1})
and~(\ref{hjm2}) are made equivalent to a single linear equation of
the form of the Schr\"odinger equation, with the Planck action
constant replaced by $\alpha$:
\begin{equation}\label{schroed}
  i\alpha\partial_t\Psi=-\frac{\alpha^2}{2m}\nabla^2\psi+V\Psi\,.
\end{equation}
We will refer to it as a Schr\"odinger--like (\Sl) equation:
clearly~(\ref{schroed}) has not the same meaning as the usual
Schr\"odinger equation; this would be true only if $\alpha=\hbar$,
while in general $\alpha$ is not an universal constant, and it is
rather a quantity characteristic of the system under consideration
(in our case the particle beam). In fact $\alpha$ turns out to be of
the order of magnitude of the beam emittance, a quantity which -- in
formal analogy with $\hbar$ -- has the dimensions of an action and
gives a measure of the position/momentum uncertainty product for the
system. Thus the SM model of our beam, as incorporated in the
phenomenological Schr\"odinger equation\refeq{schroed}, while
keeping a few features reminiscent of the QM, is in fact a deeply
different theory.

\subsection{Controlled distributions}\label{control}

We have introduced the equations that in the SM model are supposed
to describe the dynamical behavior of the beam: we now briefly sum
up a general procedure, already exploited in previous
papers~\cite{pre,jpa}, to control the dynamics of our systems. Let
us suppose that the pdf $\rho(\mathbf{r},t)$ be given all along its
time evolution: think in particular either to a stationary state, or
to an engineered evolution from some initial pdf toward a final
state with suitable characteristics. We know that the FP
equation\refeq{fp} must be satisfied, for the given $\rho$, by some
forward velocity field $\mathbf{v}_{(+)}(\mathbf{r},t)$. Since also
the equation\refeq{gradient1} must hold, we are first of all
required to find an irrotational $\mathbf{v}_{(+)}$ which satisfies
the FP equation\refeq{fp} for the given $\rho$. We then take into
account also the dynamical equation~(\ref{hjm1}): since $\rho$ and
$\mathbf{v}_{(+)}$ (and hence $f$ and $W$) are now fixed and
satisfy\refeq{fp}, the equation~(\ref{hjm1}) plays the role of a
constraint defining a controlling potential $V$ when we also take
into account the equation\refeq{phases}. We list here the potentials
associated to the three particular cases analyzed in the previous
papers.

In the 1-DIM case with given dimensionless pdf $f(x,t)$ and $a<x<b$
($a$ and $b$ can be infinite) we easily get
\begin{eqnarray}
v_{(+)}(x,t)\!\! &=&\!\!
 D\frac{\partial_x\rho(x,t)}{\rho(x,t)}-\frac{1}{\rho(x,t)}\int_a^x\partial_t\rho(x',t)\,dx'\label{1DIM}\\
V(x,t)\!\! &=&\!\! m D^2\, \partial_x^2\ln f + m D \,(\partial_t
\ln f + v_{(+)}\partial_x \ln f )\nonumber\\
&&\!\!\!\!- {m\over2}\,v_{(+)}^2 - m \int_a^x\partial_t
v_{(+)}(x',t)\,dx' + \dot\theta\label{potential1DIM}
\end{eqnarray}
For a 3-DIM system with cylindrical symmetry around the $z$-axis
(the beam axis), if we denote with $(r, \varphi, z)$ the cylindrical
coordinates, and if we suppose that $\rho(r,t)$ depends only on $r$
and $t$, and that $\mathbf{v}_{(+)}=v_{(+)}(r,t)\,\mathbf{\hat{r}}$
is radially directed with modulus depending only on $r$ and $t$, we
have
\begin{eqnarray}
v_{(+)}(r,t) \!\!&=& \!\!D\frac{\partial_r\rho(r,t)}{\rho(r,t)}
        -\frac{1}{r\rho(r,t)}\int_0^r\partial_t\rho(r',t)r'\,dr'\label{3DIM}\\
V(r,t)\!\!&=&\!\!\frac{m D^2}{r}\,
\partial_r(r\partial_r\ln f) + m D \,(\partial_t
\ln f + v_{(+)}\partial_r \ln f )\nonumber\\
&&- {m\over2}\,v_{(+)}^2 - m \int_0^r\partial_t v_{(+)}(r',t)\,dr' +
\dot\theta\label{potential3DIM}
\end{eqnarray}
Finally in the 3-DIM stationary case the pdf $\rho(\mathbf{r})$ is
independent from $t$. This greatly simplifies our formulas and, by
requiring that $\dot\theta(t)=E$ be constant, namely that
$\theta(t)=Et$, we get
\begin{eqnarray}
  \mathbf{v}_{(+)}(\mathbf{r})&=&D\frac{\nabla\rho(\mathbf{r})}{\rho(\mathbf{r})}\label{3DIMstaz} \\
  V(\mathbf{r})&=&E+2mD^2\,\frac{\nabla^2\sqrt{\rho}}{\sqrt{\rho}}\,.\label{potential3DIMstaz}
\end{eqnarray}
Of course in this context the constant $E$ will be chosen by fixing
the zero of the potential energy. Let us remark finally that in this
stationary case the phenomenological wave function\refeq{schroedpsi}
takes the form
\[
\Psi(\mathbf{r},t)=\sqrt{\rho}\,e^{-i E t/\alpha}
\]
typical of the stationary states.

\subsection{Non stationary distributions}\label{nonstationary}

In the following we will be mainly concerned with stationary
distributions, but in a few previous paper we treated also non
stationary problems. For instance, if we consider the stationary,
ground state pdf (without nodes) $\rho_0(\mathbf{r})$ of a suitable
potential, and if we calculate $\mathbf{v}_{(+)}(\mathbf{r})$ and
write down the corresponding FP equation, it is possible to show
(see the general proof in a few previous papers~\cite{fph,pla,jpa})
that, $\rho_0(\mathbf{r})$ will play the role of an attractor for
every other distribution (non extremal with respect to a stochastic
minimal action principle). If the accelerator beam is ruled by such
an equation, this would imply that the halo can not simply be wiped
out by scraping away the particles that come out of the bunch core:
in fact they simply will keep going out in the halo until the
equilibrium is reached again since the distribution
$\rho_0(\mathbf{r})$ is a stable attractor.

In a recent paper~\cite{ijmpb} we gave an estimate of the time
required for the relaxation of non extremal pdf's toward the
equilibrium distribution. This is an interesting test for our model
since this relaxation time is fixed once the form of the forward
velocity field is given; this is in turn fixed when the form of the
halo distribution is given as in the reference~\cite{prstab}, and
one could check if the estimate is in agreement with possible
observed times. In particular we estimated that in typical
conditions all the non--stationary solutions of this FP equation
will be attracted toward $\rho_0$ with a relaxation time of the
order of $\tau\approx 2m\sigma^2 /\alpha \approx 10^{-8}\div10^{-7}
\mathrm{sec}$.

A different non stationary problem also discussed in previous
papers~\cite{pre,ijmpb} consists in the analysis of some particular
time evolution of the process with the aim of finding the dynamics
that control it. For instance we studied the possible evolutions
which start from a pdf with halo and evolve toward a halo--free pdf:
this would allow us to find the dynamics that we are requested to
apply in order to achieve this result. If for simplicity the overall
process is supposed to be an Ornstein--Uhlenbeck process, the
transition pdf would be completely known and all the result can be
exactly calculated through the Chapman--Kolmogorov equation by
supposing suitable shapes for the initial and final distributions.
Then a direct application of (\ref{potential1DIM}) allows us to
calculate the control potential corresponding to this evolution. For
the sake of brevity we do not give the analytical form of this
potential and refer to the quoted papers for further details.

\section{Self-consistent equations}\label{selfconsistent}

\subsection{Space charge interaction}

In QM a system of $N$ particles is described by a wave function in a
$3N$--DIM configuration space. On the other hand in our SM scheme a
normalized $|\Psi(\mathbf{r},t)|^2$, function of only three space
coordinates $\mathbf{r}=\{x,y,z\}$, plays the role of the pdf of a
Nelson process. In a first approximation we will consider this
$N$--particle system as a pure ensemble: as a consequence we will
not introduce a $3N$--DIM configuration space, since
$N|\Psi(\mathbf{r},t)|^2\,d^3\mathbf{r}$ in the $3$--DIM space will
play the role of the number of particles in a small neighborhood of
$\mathbf{r}$. However, since our system of $N$ charged particles is
not a pure ensemble due to their mutual \EM\ interaction, in a
further \emph{mean field approximation} we will take into account
the so called \emph{space charge} effects: more precisely we will
couple our \Sl\ equation with the Maxwell equations describing both
the external and the space charge \EM\ fields, and we will get in
the end a non linear system of coupled differential equations.

In our model a single, charged particle embedded in a beam and
feeling both an external, and a space charge potential is first of
all described by a \Sl\ equation
\[
  i\alpha\partial_t\Psi(\mathbf{r},t)
  =\widehat{H}\Psi(\mathbf{r},t)\,,
\]
where $\Psi(\mathbf{r},t)$ is our wave function, $\alpha$ a
coefficient with the dimensions of an action which is a constant
depending on the beam characteristics, and $\widehat{H}$ is a
suitable Hamiltonian operator. If $\Psi$ is properly normalized and
if $N$ is the number of particles with individual charge $q_0$, the
space charge density and the electrical current density are
\begin{eqnarray}
\rho_{sc}(\mathbf{r},t)&=&Nq_0|\Psi(\mathbf{r},t)|^2\,,\label{defdensity}\\
 \mathbf{j}_{sc}(\mathbf{r},t)&=&Nq_0\frac{\alpha}{m}\,
   \Im\,\{\Psi^*(\mathbf{r},t)\mathbf{\nabla}\Psi(\mathbf{r},t)\}\,.\label{defcurrent}
\end{eqnarray}
Hence our particles in the beam will feel both an electrical and a
magnetic interaction and we will be obliged to couple the \Sl\
equation with the equations of the vector and scalar potentials
associated to this electro--magnetic field.

The \EM\ potentials $(\mathbf{A}_{sc},\Phi_{sc})$ of the space
charge fields obeying the gauge condition
\begin{equation}\label{gauge}
  \nabla\cdot\mathbf{A}_{sc}(\mathbf{r},t)+\frac{1}{c^2}\,\partial_t\Phi_{sc}(\mathbf{r},t)=0\,,
\end{equation}
must satisfy the wave equations
\begin{eqnarray}
 \nabla^2\mathbf{A}_{sc}(\mathbf{r},t)-\frac{1}{c^2}\,\partial_t^2\mathbf{A}_{sc}(\mathbf{r},t)
   &=&-\mu_0\mathbf{j}_{sc}(\mathbf{r},t)\label{waveqA}\\
 \nabla^2\Phi_{sc}(\mathbf{r},t)-\frac{1}{c^2}\,\partial_t^2\Phi_{sc}(\mathbf{r},t)
   &=&-\frac{\rho_{sc}(\mathbf{r},t)}{\epsilon_0}\label{waveqFi}
\end{eqnarray}
On the other hand, for our particle in the beam the \EM\ field is
the superposition of the space charge potential
$(\mathbf{A}_{sc},\Phi_{sc})$, and of the external potentials
$(\mathbf{A}_e,\Phi_e)$. Hence (see for example \cite{landau2},
chapter XV) our \Sl\ equation takes the form
\begin{eqnarray}\label{schroedinger}
  i\alpha\partial_t\Psi&=&\frac{1}{2m}
  \left[i\alpha\nabla-\frac{q_0}{c}(\mathbf{A}_{sc}+\mathbf{A}_e)\right]^2\Psi\nonumber\\
  && \qquad\qquad\qquad\qquad  +\,q_0(\Phi_{sc}+\Phi_e)\Psi
\end{eqnarray}
It is apparent now that\refeq{gauge},\refeq{waveqA},\refeq{waveqFi}
and\refeq{schroedinger} constitute a self--consistent system of non
linear differential equations for the fields $\Psi$,
$\mathbf{A}_{sc}$ and $\Phi_{sc}$ coupled through\refeq{defdensity}
and\refeq{defcurrent}.

If we then consider stationary wave functions
\begin{equation}\label{stationarywf}
\Psi(\mathbf{r},t)=\psi(\mathbf{r})\,e^{-iEt/\alpha}
\end{equation}
where $E$ is the energy of the particle, and take $\mathbf{A}_e=0$
for the external interaction, passing to the potential energies
\[
V_e(\mathbf{r})=q_0\Phi_e(\mathbf{r})\,,\qquad
V_{sc}(\mathbf{r})=q_0\Phi_{sc}(\mathbf{r})\,,
\]
our system is reduced to only two coupled, non linear equations for
the pair $(\psi,V_{sc})$, namely
\begin{eqnarray}
 \frac{\alpha^2}{2m}\nabla^2\psi&=&(V_e+V_{sc}-E)\psi\,,\label{selfconsist1}\\
 \nabla^2V_{sc}&=&
            -\frac{Nq_0^2}{\epsilon_0}\,|\psi|^2\label{selfconsist2}
\end{eqnarray}

\subsection{Cylindrical symmetry}

We suppose now that the longitudinal motion along the $z$--axis is
both decoupled from the transverse motion in the $x,y$--plane, and
free with constant momentum $p_z$, and velocity $b_z=b_0\gg
b_x,b_y$. Moreover we suppose that the beam particles will be
confined in a cylindrical packet of length $L$, so that by the
imposing periodic boundary conditions we will quantize the
longitudinal momentum
\begin{equation*}
  p_z=\frac{2k\pi\alpha}{L}\,,\qquad\quad k=0,\pm1,\pm2,\ldots
\end{equation*}
As a consequence our wave functions will take the form
\begin{equation}\label{stationarystate}
\psi(\mathbf{r})=\chi(x,y)\,\frac{e^{ip_zz/\alpha}}{\sqrt{L}}
\end{equation}
and our equations\refeq{selfconsist1} and\refeq{selfconsist2} become
\begin{eqnarray}
 \frac{\alpha^2}{2m}(\partial^2_x+\partial^2_y)\chi\!\!&=&\!\!\left(V_e+V_{sc}-E_T\right)\,\chi\label{selfconstat1}\\
 (\partial^2_x+\partial^2_y)V_{sc}\!\!&=&\!\!-\frac{Nq_0^2}{L\epsilon_0}\,|\chi|^2=
            -\frac{\mathcal{N}q_0^2}{\epsilon_0}\,|\chi|^2\label{selfconstat2}
\end{eqnarray}
where $\mathcal{N}=N/L$ is the number of particles per unit length,
and $E_T=E-p_z^2/2m$ is the energy of the transverse motion. If
finally our system has a cylindrical symmetry around the $z$ axis,
namely if -- in the cylindrical coordinate system $\{r,\varphi,z\}$
($r^2=x^2+y^2$) -- our potentials depend only on $r$, then we can
separate the variables with $\chi(x,y)=u(r)\Phi(\varphi)$, the
angular eigenfunctions are
\begin{equation}\label{angular}
  \Phi_\ell(\varphi)=\frac{e^{i\ell\varphi}}{\sqrt{2\pi}}\,,\qquad\quad\ell=0,\pm1,\pm2,\ldots
\end{equation}
and for $\ell=0$ the equations become
\begin{eqnarray}
 \frac{\alpha^2}{2m}\left(u''+\frac{u'}{r}\right)&=&
      \left(V_e+V_{sc}-E_T\right)u\label{cylindricalsc1}\\
       V''_{sc}+ \frac{V'_{sc}}{r}&=&-\frac{\mathcal{N}q_0^2}{2\pi\epsilon_0}\,u^2\label{cylindricalsc2}
\end{eqnarray}
with the following radial normalization
\[
\int_0^{+\infty}ru^2(r)\,dr=1\,.
\]
Remark that now we are reduced to a system of \emph{ordinary}
differential equations.

\subsection{Dimensionless formulation}\label{dimlessform}

To eliminate the physical dimensions one introduces two quantities
$\eta$ and $\lambda$ which are respectively an energy and a length.
Then, by means of the dimensionless quantities
\begin{eqnarray*}
  s\!\!&=&\!\!\frac{r}{\lambda}\,,\qquad
  \beta=\frac{E_T}{\eta}\,,\qquad
  \xi=\frac{\mathcal{N}q_0^2}{2\pi\epsilon_0\eta}\;\;\;(\mbox{\emph{perveance}})\\
  &&\qquad\qquad w(s)=\lambda\, u(\lambda s) \\
  &&v(s)=\frac{V_{sc}(\lambda s)}{\eta}\,,\qquad
  v_e(s)=\frac{V_{e}(\lambda s)}{\eta}
\end{eqnarray*}
the equations\refeq{cylindricalsc1} and\refeq{cylindricalsc2} take
the form
\begin{eqnarray}
  s \,w''(s)+w'(s)&=&\left[v_e(s)+v(s)-\beta\right]s\, w(s)\label{cylindrscdl1} \\
  s \,v''(s)+v'(s)&=&-\,\xi\, s\, w^2(s)\label{cylindrscdl2}
\end{eqnarray}
The usual choice for the dimensional constants is
\begin{equation}\label{dimensionalconstants}
  \eta=mb_0^2\,,\qquad\quad
  \lambda=\frac{\alpha}{mb_0\sqrt{2}}\,,
\end{equation}
where $b_0$ is the longitudinal velocity of the beam. We can now
look at our equations in two different ways. First of all we can
suppose that $v_e$ is a given external potential: in this case our
aim is to solve the equations for the two unknowns $w$ (radial
particle distribution) and $v$ (space charge potential energy).
However in general no simple analytical solution of this problem is
at present available for the usual forms of the external potential
$v_e$: there are not even solutions playing the same role played by
the Kapchinskij--Vladimirskij (KV) distribution in the usual models.
This phase space distribution -- which is simple and self-consistent
in the usual dynamical models -- leads to an uniform transverse
space distribution of the beam, and is a stationary solution of the
Vlasov equation with a harmonic potential. Moreover its space charge
potential calculated from the Poisson equation is still harmonic.
Instead in the SM model the uniform distributions are not solutions
of the stationary Schr\"odinger equation, and we know no simple
stationary distribution connected to the harmonic potential as the
KV. Even the gaussian distributions -- later discussed in this paper
-- can not play the same role: they are solutions connected with an
external harmonic potential, but their space charge potential
calculated from the Poisson equation is not harmonic.

Alternatively we can assume as known a given distribution $w$, and
solve our equations to find both the external and the space charge
self--consistent potential energies $v_e$ and $v$. In this second
form the problem is more simple, and analytical solutions are
available. We adopted the first standpoint in a few previous
papers~\cite{prstab} where we numerically solved the
equations\refeq{cylindrscdl1} and\refeq{cylindrscdl2}; here we will
rather elaborate a few new ideas about the second one. To this end
it is important to remark that the space charge potential energy
\begin{equation}\label{poissonsolution}
  v(s)=-\xi\int_0^s\frac{dy}{y}\int_0^yxw^2(x)\,dx
\end{equation}
is always a solution of the Poisson equation\refeq{cylindrscdl2}
satisfying the conditions $v(0^+)=v'(0^+)=0$. On the other hand, by
substituting\refeq{poissonsolution} in the first
equation\refeq{cylindrscdl1} we readily obtain also the
self--consistent form of the external potential energy
\begin{eqnarray}
  v_e(s)&=&v_0(s) +
  \xi\int_0^s\frac{dy}{y}\int_0^yxw^2(x)\,dx\,,\label{integrodifferential}\\
    v_0(s)&=&\frac{w''(s)}{w(s)}+\frac{1}{s}\frac{w'(s)}{w(s)}+\beta\label{potentialnosc}
\end{eqnarray}
where $v_0(s)$ is the potential that we would have without space
charge ($\xi=0$), while the second part in the external
potential\refeq{integrodifferential} exactly compensate for the
space charge potential.

\subsection{Constant focusing}

Let us suppose now that the transverse external potential $V_e(r)$
is a cylindrically symmetric, harmonic potential with a proper
frequency $\omega$ (\emph{constant focusing}), and let we also
introduce the characteristic length
\[
\sigma^2=\frac{\alpha}{2m\omega}
\]
which will represents a measure of the transverse dispersion of the
beam. In cylindrical coordinates $\{r,\varphi\}$ in the transverse
plane our potential energy is
\begin{equation}\label{cylindrHO}
V_e(r)=\frac{m\omega^2}{2}\,r^2=\frac{\alpha^2}{8m\sigma^4}\,r^2
\end{equation}
so that the corresponding 2--DIM \Sl\ equation \emph{without} space
charge (zero perveance) would have as lowest eigenvalue
$E_0=\alpha\omega$, and as ground state wave function
\begin{equation}\label{unpertground}
\chi_{00}(r,\varphi)=\frac{u_0(r)}{\sqrt{2\pi}}
=\frac{e^{-r^2/4\sigma^2}}{\sigma\sqrt{2\pi}}\,.
\end{equation}
Of course the self--consistent solution would be different if there
is a space charge (non zero perveance). To find this solution one
introduces the so called \emph{phase advance}
\begin{equation*}
  \frac{1}{\lambda_0}=\frac{\omega}{b_0}=\frac{\alpha}{2mb_0\sigma^2}
\end{equation*}
($\lambda_0$ is a length) and, with the
constants\refeq{dimensionalconstants}, the dimensionless form of the
harmonic potential\refeq{cylindrHO}
\begin{eqnarray*}
  v_e(s)&=&\frac{V_e(r)}{mb_0^2}=\frac{\omega^2}{b_0^2}\,r^2=\frac{r^2}{2\lambda^2_0}
        =\frac{\alpha^2}{4\lambda_0^2\,m^2\,b_0^2}\,s^2=\gamma^2\,s^2\\
  \gamma&=&\frac{\alpha}{2\lambda_0mb_0}=\frac{\alpha\omega}{2mb_0^2}=\frac{\sigma^2}{\lambda_0^2}\,.
\end{eqnarray*}
As a consequence the equations\refeq{cylindrscdl1}
and\refeq{cylindrscdl2} become
\begin{eqnarray}
  s \,w''(s)+w'(s)&=&\left[\gamma^2\,s^2+v(s)-\beta\right]s\, w(s) \label{cylindrscdlHO1}\\
  s \,v''(s)+v'(s)&=&-\,\xi\, s\, w^2(s) \label{cylindrscdlHO2}
\end{eqnarray}
These equations are now a coupled, non linear system which must be
\emph{numerically} solved since we do not know simple
self--consistent solutions of the form of the KV distribution. In
reference~\cite{prstab} we extensively analyzed these numerical
solutions and we refer to this paper for details. In fact
in~\cite{prstab} there was a small difference with respect to what
has been presented here. The form of the equations to solve is the
same, but the dimensionless formulation was achieved by means of two
numerical constants different from\refeq{dimensionalconstants} and
drawn from the characteristics of the transverse harmonic oscillator
force:
\begin{equation}
  \eta=\frac{\alpha^2}{4m\sigma^2}=\frac{\alpha\omega}{2}\,,\qquad\lambda=\sigma\sqrt{2}
\end{equation}
Then the dimensionless quantities have a different numerical value
and the dimensionless equations\refeq{cylindricalsc1}
and\refeq{cylindricalsc2} take the form
\begin{eqnarray}
  {s}\, {w}''({s})+{w}'({s})
  &=&[{s}^2+{v}({s})-{\beta}]\,{s}\, {w}({s})\label{adimsc1}\\
  {s}\, {v}''({s})+{v}'({s})&=&-\,{\xi}\,{s}\,
  {w}^2({s}) \label{adimsc2}
\end{eqnarray}
since now $\gamma=1$. In any case the
equations\refeq{cylindrscdlHO1} and\refeq{cylindrscdlHO2} can easily
be turned into the equations\refeq{adimsc1} and\refeq{adimsc2}, and
\emph{vice versa}, by means of simple transformations through the
parameter $\gamma$ which turns out to be at the same time the ratio
of the energy constants, and that of the squared length constants.
As a consequence in the following we will always use the
system\refeq{adimsc1},\refeq{adimsc2}, with the advantage of simply
putting $\gamma=1$ in the model.

\section{Self--consistent potentials}\label{selfconspot}

\subsection{Gaussian transverse distributions}\label{gaussdistr}

In the SM model it is possible to numerically integrate the
Schr\"odinger--Poisson system\refeq{cylindrscdl1}
and\refeq{cylindrscdl2} with a given external potential and
calculate the self--consistent distributions and their space charge
potentials~\cite{prstab}. On the other hand, if we fix a particular
distribution, it is always possible to exactly calculate from these
equations the external and space charge potential giving rise to
that distribution. When we adopt this second alternative approach
and we take as given the form of the distribution $w(s)$, the
unknowns in the equations\refeq{cylindrscdl1}
and\refeq{cylindrscdl2} are the two potential energies $v(s)$ and
$v_e(s)$. In this case we only need to calculate the
expressions\refeq{poissonsolution} and\refeq{integrodifferential} in
terms of the given distribution $w(s)$. Of course if we take an
arbitrary $w(s)$ we will not get any simple and meaningful form for
the external potential $v_e(s)$; and on the other hand to guess the
right form of $w(s)$ giving rise, for instance, exactly to a
harmonic potential\refeq{cylindrHO} as external potential would be
tantamount to solve\refeq{integrodifferential} as an
integro-differential equation for a given external potential.
However in a few explicit cases the results are quite simple and
interesting.

\begin{figure}
\begin{center}
\includegraphics*[width=8.0cm]{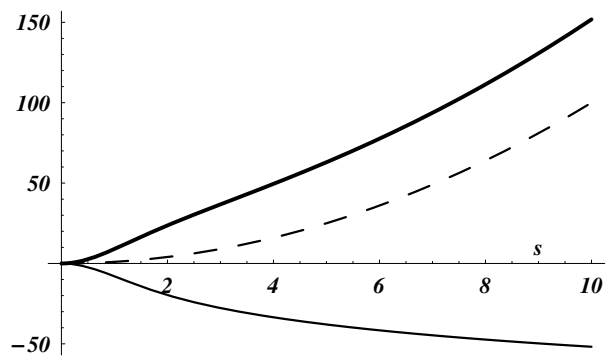}
\end{center}
\caption{The dimensionless potentials $v(s)$ (thin line),
$v_0(s)=s^2$ (dashed line) and $v_e(s)$ (thick line). They reproduce
respectively equations\refeq{schopot1},\refeq{schopot3}
and\refeq{schopot2} for $\xi=20$ (see reference~\cite{prstab} for
this value). When the external potential is $v_e(s)$ the
self--consistent wave function coincides with that of a simple
harmonic oscillator for zero
perveance\refeq{dlstationary1}.}\label{potentials}
\end{figure}

Let us take as first example of a stationary wave function that of
the ground state $u_0(r)$ of the harmonic oscillator with zero
perveance given in\refeq{unpertground}. Its dimensionless
representation is:
\begin{equation}
  w(s)=\sqrt{2}\,e^{-s^2/2}\,,\qquad\beta=2\quad(E_T=\alpha\omega)\,;\label{dlstationary1}
  \end{equation}
which is also apparently normalized. We now want to calculate both
the external and the space charge potentials that
produce\refeq{dlstationary1} as stationary wave function
for\refeq{adimsc1} and\refeq{adimsc2}.
From\refeq{poissonsolution},\refeq{integrodifferential}
and\refeq{dlstationary1} we then have
\begin{eqnarray*}
  \frac{w''(s)}{w(s)}+\frac{1}{s}\,\frac{w'(s)}{w(s)}+\beta&=&v_0(s)=s^2 \\
  \int_0^s\frac{dy}{y}\int_0^yw^2(x)x\,dx&=&\frac{1}{2}\left[\log(s^2)+\mathbb{C}-\mathrm{Ei}(-s^2)\right]
\end{eqnarray*}
where $\mathbb{C}\approx0.577$ is the Euler constant and
\[
\mathrm{Ei}(x)=\int_{-\infty}^x\frac{e^t}{t}\,dt\,,\qquad\quad x<0
\]
is the exponential--integral function, and hence we immediately get
(see also FIG.\myref{potentials})
\begin{eqnarray}
  v(s)&=&-\frac{\xi}{2}\left[\log(s^2)+\mathbb{C}-\mathrm{Ei}(-s^2)\right]\label{schopot1} \\
  v_0(s)&=&s^2\label{schopot3}\\
  v_e(s)&=&s^2+\frac{\xi}{2}\left[\log(s^2)+\mathbb{C}-\mathrm{Ei}(-s^2)\right]\label{schopot2}
\end{eqnarray}
In a sense the meaning of the
equations\refeq{poissonsolution},\refeq{integrodifferential}
and\refeq{potentialnosc} is rather simple: if we want to get a
self-consistent distribution which coincides with a solution of the
\Sl\ equation for a given zero perveance potential, the simplest way
it is to calculate the space charge potential for this \emph{frozen}
distribution through the Poisson equation, and then compensate the
external potential exactly for that. This is what we did in our
example where the gaussian solution is the fundamental state of a
harmonic oscillator: we finally got a total potential which is
$v_0(s)=v(s)+v_e(s)= s^2$ (namely that of a simple harmonic
oscillator), and an energy value which coincides with the first
eigenvalue. In other words, if you want a gaussian transverse
distribution you should not simply turn on a bare harmonic potential
$s^2$: you should rather teleologically compensate for the space
charge by using the potential $v_e(s)$.

\subsection{Student transverse distributions}\label{studtransvdist}

If the halo consists in the fact that large deviations from the beam
axis are possible, a new idea is to suppose that the the stationary
transverse distribution is different from the gaussian
distribution\refeq{dlstationary1} introduced in the
Section\myref{gaussdistr}. To this end we will introduce in the
following a family of distributions which decay with the distance
from the axis only with a power law.

\begin{figure}
\begin{center}
\includegraphics*[width=8.0cm]{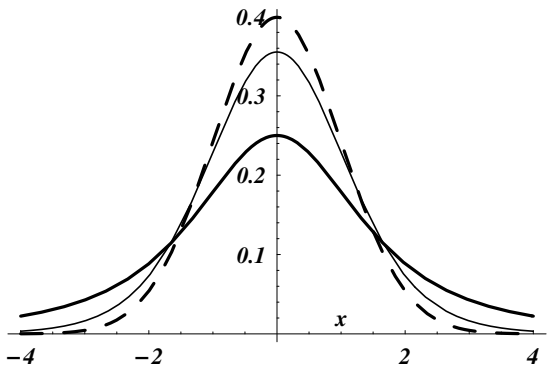}
\end{center}
\caption{The Gauss pdf $\mathcal{N}(0,1)$ (dashed line) compared
with the $\Sigma(2,2)$ (thick line) and the $\Sigma(10,12)$ (thin
line). The flexes of the three curves coincide. Apparently the tails
of the Student laws are much longer.}\label{GaussStud}
\end{figure}

Let us consider the following family of univariate, two--parameters
probability laws $\Sigma(\nu,a^2)$ characterized by the following
pdf's
\begin{equation}\label{1Dstudentpdf}
    f(x)=\frac{\Gamma\left(\frac{\nu+1}{2}\right)}{\Gamma\left(\frac{1}{2}\right)\Gamma\left(\frac{\nu}{2}\right)}\,
    \frac{a^\nu}{(x^2+a^2)^{\frac{\nu+1}{2}}}
\end{equation}
which apparently are symmetric functions with the mode in $x=0$ and
two flexes in $x=\pm a/\sqrt{\nu+2}$. All these laws are centered at
the median. In particular $a$ plays just the role of a scale
parameter, while $\nu$ rules the power decay of the tails: for large
$x$ the tails go as $x^{-(\nu+1)}$ with $\nu+1>1$. For a comparison
with a Gauss law $\mathcal{N}(0,\sigma^2)$ see
FIG.\myref{GaussStud}. Remark that when $\nu$ grows larger and
larger, the difference between the two pdf's becomes smaller and
smaller.
It is typical of the laws $\Sigma(\nu,a^2)$ that they have (finite)
momenta of order $k$ only if the condition $k<\nu$ is verified;
hence for $\nu\leq2$ there is no variance, while for $\nu\leq1$ not
even the expectation is defined. On the other hand when $\nu>2$ the
variance of $\Sigma(\nu,a^2)$ exists and is
\begin{equation}\label{StudentVar}
    \sigma^2=\frac{a^{2}}{\nu-2}.
\end{equation}
It will be useful to remark that the laws  $\Sigma(1,a^2)$ are the
well--known Cauchy laws $\mathcal{C}(a)$ with pdf
\begin{equation*}
    f(x)=\frac{1}{\pi}\frac{a}{x^{2}+a^{2}}\,,
\end{equation*}
while the laws  $\Sigma(n,n)$ with $n=1,2,\ldots$ are the classical
$t$--Student laws $\mathcal{S}(n)$ with pdf
\begin{equation*}
    f(x)=\frac{\Gamma\left(\frac{n+1}{2}\right)}{\sqrt{\pi}\,\Gamma\left(\frac{n}{2}\right)}(n+x^{2})^{-\frac{n+1}{2}}\,.
\end{equation*}
We will then refer to $\Sigma(\nu,a^2)$ as generalized Student laws
since they are just Student laws with a continuous parameter $\nu>0$
and a scale parameter $a$. For $\nu>2$ variances exist and we are
then entitled to standardize our laws: indeed from\refeq{StudentVar}
every $\Sigma(\nu,(\nu-2)\sigma^2)$ with $a^2=(\nu-2)\sigma^2$ has
variance $\sigma^2$, and the standard (with unit variance)
generalized Student laws are $\Sigma(\nu,\nu-2)$.

\begin{figure}
\begin{center}
\includegraphics*[width=8.0cm]{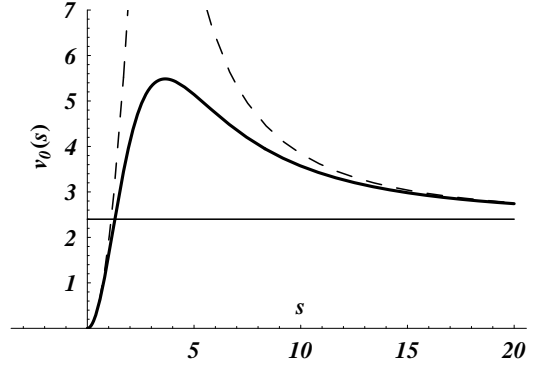}
\end{center}
\caption{The control potential $v_0(s)$\refeq{3Dstudent0} for a
Student transverse distribution $\Sigma_2(22,20\sigma^{2})$. Also
displayed are the value of $\beta=2.4$ (the limit value of $v_0$ for
large $s$, thin line) and the behaviors for small and large
$s$\refeq{zerobehav} and\refeq{infbehav} (dashed
lines).}\label{studentalone}
\end{figure}

\begin{figure}
\begin{center}
\includegraphics*[width=8.0cm]{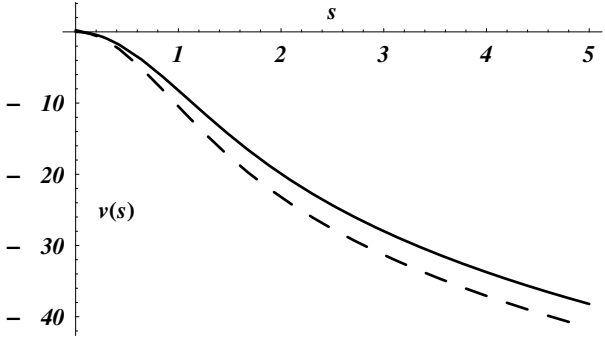}
\end{center}
\caption{The space charge potentials $v(s)$\refeq{3Dstudentsc}
and\refeq{schopot1} respectively for a Student (solid line)
transverse distribution $\Sigma_2(22,20\sigma^{2})$, and for a Gauss
(dashed line) distribution. The dimensionless perveance here is
$\xi=20$.}\label{studentSC}
\end{figure}

\begin{figure}
\begin{center}
\includegraphics*[width=8.5cm]{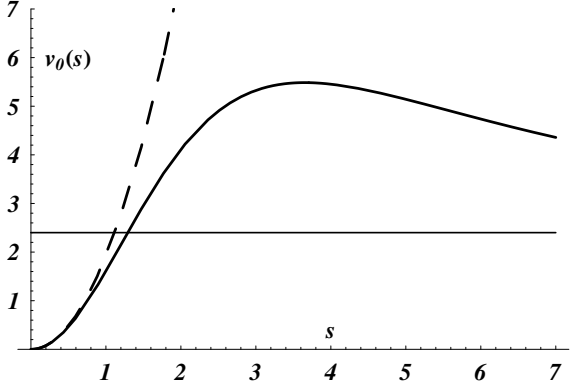}
\end{center}
\caption{The control potential $v_0(s)$\refeq{3Dstudent0} of a
Student $\Sigma_2(22,20\sigma^{2})$ (solid line; see
FIG.\myref{studentalone}) is here compared with that of a Gauss
distribution (dashed line) which shows the same behavior near the
beam axis.}\label{studentpot}
\end{figure}

\begin{figure}
\begin{center}
\includegraphics*[width=8.0cm]{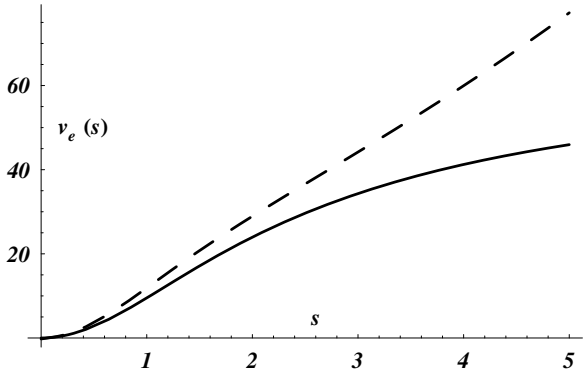}
\end{center}
\caption{The total external potential $v_e(s)$\refeq{3Dstudentext}
that should be applied to get a stationary Student transverse
distribution $\Sigma_2(22,20\sigma^{2})$ (solid line), compared with
that\refeq{schopot2} needed for a Gauss distribution (dashed
line).}\label{studentext}
\end{figure}

In order to describe the beam we will also introduce the bivariate,
circularly symmetric Student laws $\Sigma_2(\nu,a^2)$ with pdf
\begin{equation}\label{2Dstudentpdf}
    f(x,y)=\frac{\nu}{2\pi}\frac{a^{\nu}}{(x^{2}+y^{2}+a^{2})^{\frac{\nu+2}{2}}}\,.
\end{equation}
Its marginal laws are both $\Sigma(\nu,a^2)$ and non--correlated,
albeit not independent (as in the case of the circularly symmetric
gaussian bivariate laws). The total beam distribution will then be
\begin{equation}\label{3Dbeamstudent}
    \rho(x,y,z)=\frac{1}{2\pi L}\,\frac{\nu
    a^{\nu}}{(x^{2}+y^{2}+a^{2})^{\frac{\nu+2}{2}}}\,H\left(\frac{L}{2}-|z|\right)
\end{equation}
where $H(z)$ is the Heaviside function. In the description of a beam
in an accelerator it is realistic to suppose that the transverse
distribution is endowed with a finite variance. Hence we will look
for distributions\refeq{3Dbeamstudent} with $\nu>2$. On the other
hand this will correspond to suppose that in our model the
transverse Student laws should not be radically different from a
Gaussian: in fact the halo is in some sense an effect which is small
when compared with the total beam. From this standpoint the family
of laws $\Sigma(\nu,a^2)$ has also the advantage that we can fine
tune the parameters $\nu,a$ in order to get the right distance from
the gaussian laws (this would not be possible if we adopted stable
laws; see subsequent Section\myref{studentid}). With this hypothesis
in mind we will limit our present considerations to the case $\nu>2$
so that the transverse marginals of\refeq{3Dbeamstudent} will have a
finite variance $\sigma^2$. Then from\refeq{StudentVar} we choose
$a^2=(\nu-2)\sigma^2$ and write\refeq{3Dbeamstudent} as
\begin{equation}\label{3DbeamstudentVar}
    \rho(x,y,z)=\frac{\nu}{2\pi L}\,\frac{[(\nu-2)\sigma^2]^{\frac{\nu}{2}}}
    {[x^{2}+y^{2}+(\nu-2)\sigma^{2}]^{\frac{\nu+2}{2}}}\,H\left(\frac{L}{2}-|z|\right)
\end{equation}
Passing to cylindrical random variables we then have
\begin{equation*}
    \rho(r,\varphi,z)=r\,\frac{\nu}{2\pi L}\,\frac{[(\nu-2)\sigma^2]^{\frac{\nu}{2}}}
    {[r^{2}+(\nu-2)\sigma^{2}]^{\frac{\nu+2}{2}}}\,H\left(\frac{L}{2}-|z|\right)
\end{equation*}
namely
\begin{eqnarray*}
    \rho(r,\varphi,z)&=&\frac{1}{\sigma\sqrt{2}}\,\frac{r}{\sigma\sqrt{2}}\,\frac{2\nu}{\nu-2}\\
    &&\qquad\times
    \left[1+\frac{r^{2}}{(\nu-2)\sigma^2}\right]^{-\frac{\nu+2}{2}}\,\frac{H\left(\frac{L}{2}-|z|\right)}{2\pi L}
\end{eqnarray*}
so that finally with the shorthand notation
\begin{equation}\label{shorthand}
    z=\frac{s\sqrt{2}}{\sqrt{\nu-2}}
\end{equation}
the dimensionless, normalized radial distribution is
\begin{equation}\label{dlradialstudent}
    w^2(s)=\frac{2\nu}{\nu-2}\frac{1}{(1+z^2)^{\frac{\nu+2}{2}}}.
\end{equation}
Here we adopt the dimensional constants
\begin{equation}
  \eta=\frac{\alpha^2}{4m\sigma^2}\,,\qquad\lambda=\sigma\sqrt{2}
\end{equation}
where $\sigma^2$ is the variance of our Student laws. We can now use
the relations\refeq{poissonsolution},\refeq{integrodifferential}
and\refeq{potentialnosc} in order to get the potentials which
have\refeq{3DbeamstudentVar} as stationary distribution: first of
all the space charge potential produced by\refeq{3DbeamstudentVar}
has the form
\begin{eqnarray}\label{3Dstudentsc}
  v(s) \!\!&=&\!\!
  -\frac{\xi}{2}\left[
             \frac{2z^{-\nu}}{\nu}\,\hyperg\left(\frac{\nu}{2}\,,\frac{\nu}{2}\,;\frac{\nu+2}{2}\,;-\frac{1}{z^2}\right)\right.\nonumber \\
   && \left.\qquad\quad+\log z^2+\mathbb{C}+\psi\left(\frac{\nu}{2}\right)\right]
\end{eqnarray}
where $\hyperg(a,b;c;w)$ is a hypergeometric function and
$\psi(w)=\Gamma'(w)/\Gamma(w)$ is the logarithmic derivative of the
Euler Gamma function (digamma function). On the other hand, by
choosing $\beta= 2+\frac{8}{\nu-2}$ to put the potential energies to
zero in the origin, we get the control potential for zero perveance
\begin{equation}\label{3Dstudent0}
    v_0(s)=\frac{\nu+2}{\nu-2}\,
    \frac{z^2(4z^2+\nu+10)}{2(1+z^2)^2}
\end{equation}
and hence the external potential required to keep a transverse
student distribution $\Sigma_2(\nu,(\nu-2)\sigma^2)$ with a given
variance $\sigma^2$ is
\begin{eqnarray}\label{3Dstudentext}
  v_e(s) &=& \frac{\nu+2}{\nu-2}\,
    \frac{z^2(4z^2+\nu+10)}{2(1+z^2)^2} \nonumber \\
  && \qquad+\frac{\xi}{2}\left[
             \frac{2z^{-\nu}}{\nu}\,\hyperg\left(\frac{\nu}{2}\,,\frac{\nu}{2}\,;\frac{\nu+2}{2}\,;-\frac{1}{z^2}\right)\right.\nonumber \\
   && \left.\qquad\qquad+\log z^2+\mathbb{C}+\psi\left(\frac{\nu}{2}\right)\right]
\end{eqnarray}
Formulas\refeq{3Dstudentsc},\refeq{3Dstudent0}
and\refeq{3Dstudentext} give the self--consistent potentials
associated with the beam distribution\refeq{3DbeamstudentVar} which
is transversally a Student $\Sigma_2(\nu,(\nu-2)\sigma^2)$. In the
FIG.\myref{studentalone} we can see an example of the control
potential $v_0(s)$ for a particular value of the parameter $\nu$,
together with its limit behaviors
\begin{eqnarray}
  v_0(s) &\sim& \frac{(\nu+2)(\nu+10)}{(\nu-2)^{2}}\,s^{2}\,,\qquad\;\;(s\rightarrow0^+)\label{zerobehav}\\
  v_0(s) &\sim& \frac{(\nu+2)^2}{4s^{2}}+2+\frac{8}{\nu-2}\,,\quad\,(s\rightarrow+\infty)\label{infbehav}
\end{eqnarray}
Now this results must be compared with the similar
results\refeq{schopot1},\refeq{schopot3} and\refeq{schopot2}
associated to a transversally gaussian distribution. We will choose
the gaussian parameters in such a way that the behavior near the
beam axis be similar to\refeq{zerobehav}, namely (with
$\beta=2\gamma$)
\begin{equation*}
    w(s)=\sqrt{2\gamma}\,e^{-2\gamma
    s^2/2},\qquad\gamma^2=\frac{(\nu+2)(\nu+10)}{(\nu-2)^{2}}.
\end{equation*}
First of all in the FIG.\myref{studentSC} we compare the space
charge potential produced by both a Student and Gauss transverse
distribution: remark as for the chosen parameter values
($\nu=22,\,\xi=20$) the two potentials look particularly similar. In
fact, given the asymptotic behavior of the hypergeometric function
in\refeq{3Dstudentsc} and of the exponential integral
in\refeq{schopot1}, for $s\rightarrow+\infty$ both potentials behave
as $-\xi\log s$. On the other hand we immediately see from
FIG.\myref{studentpot} that the control potentials for zero
perveance $v_0(s)$ behave differently when we move away from the
beam axis; beyond a distance of about $r\simeq2\sigma$ the two
curves are different: while in the Gaussian case the potential
diverges as $s^2$, in the Student case it goes to the constant value
$\beta$ as quickly as $s^{-2}$. Of course this difference fades away
when $\nu$ grows larger and larger; that points to the fact that the
principal difference between the two cases can be confined in a
region that can be made as far removed from the beam core as we want
by a suitable choice of $\nu$. Finally in the FIG.\myref{studentext}
we compare the total external potentials needed to keep the
transverse beam respectively in a Student and in a Gauss
distribution. We then see that for large $s$ (far away from the beam
core, while in the Gauss case the total external potential grows
with $s$ as $s^2+\xi\log s$, in the Student case this potential only
grows as $\xi\log s$. In any case, even if the potential near the
beam axis is harmonic, deviations from this behavior in a region
removed form the core can produce a deformation of the distribution
from the gaussian to the Student.

\subsection{Estimating the emittance}\label{esimatingemittance}

If $u(r)$ is a self--consistent, cylindrically symmetric solution
of\refeq{cylindricalsc1} and\refeq{cylindricalsc2} the position
probability density $\rho(r,\varphi,z)$ in cylindrical coordinates
will have the form
\begin{equation*}
  \rho(r,\varphi,z)=\frac{1}{2\pi L}
  \begin{cases}
    u^2(r) & 0\leq\varphi<2\pi, \quad -\frac{L}{2}\leq z\leq \frac{L}{2}, \\
    0 & \text{otherwise}.
  \end{cases}
\end{equation*}
In order to estimate the emittance we need to calculate mean values
of positions $x$ and momenta $p_x$ along one transverse direction,
but we should remember that in SM we have neither a distribution in
the phase space, nor an operator formalism. The momentum and its
distribution should then be recovered from the velocity
fields\refeq{osmotic} and\refeq{gradient} where -- since we are
dealing with stationary states with $\mathbf{v}=0$ -- only the
osmotic part is non zero so that
\begin{equation*}
  \mathbf{p}=\alpha\,\frac{\nabla\rho}{\rho}
=2\alpha\,\frac{\nabla u(r)}{u(r)}\,.
\end{equation*}
By supposing now to choose the $\nu$ of our Student laws so that the
following integrals exist, we then have
\begin{eqnarray*}
  \langle x\rangle&=&\int_0^L\frac{dz}{L}\int_0^{2\pi}\frac{\cos\varphi}{2\pi}\,d\varphi\int_0^{+\infty}r^2u^2(r)\,dr=0\\
  \langle
  x^2\rangle&=&\int_0^L\frac{dz}{L}\int_0^{2\pi}\frac{\cos^2\varphi}{2\pi}\,d\varphi\int_0^{+\infty}r^3u^2(r)\,dr\\
                     &=&\frac{1}{2}\int_0^{+\infty}r^3u^2(r)\,dr\\
  \langle p_x\rangle&=&2\alpha\int_0^L\frac{dz}{L}\int_0^{2\pi}\frac{\cos\varphi}{2\pi}\,d\varphi
                           \int_0^{+\infty}r u(r)u'(r)\,dr=0\\
  \langle p_x^2\rangle&=&4\alpha^2\int_0^L\frac{dz}{L}\int_0^{2\pi}\frac{\cos^2\varphi}{2\pi}\,d\varphi\int_0^{+\infty}r
  u'\,^2(r)\,dr\\
                     &=&2\alpha^2\int_0^{+\infty}r u'\,^2(r)\,dr
\end{eqnarray*}
so that the standard deviations (uncertainties) are
\begin{eqnarray}
  \Delta x&=&\sqrt{\langle x^2\rangle}
                  =\sqrt{\frac{1}{2}\int_0^{+\infty}r^3u^2(r)\,dr}\label{deltax}\\
  \Delta p_x&=&\sqrt{\langle p_x^2\rangle}
                  =\sqrt{2\alpha^2\int_0^{+\infty}r u'\,^2(r)\,dr}\label{deltap}
\end{eqnarray}
and the position--momentum covariance is
\begin{eqnarray*}
  C&=&\langle x p_x\rangle-\langle x\rangle\langle
                               p_x\rangle=\langle x p_x\rangle\\
           &=&2\alpha\int_0^L\frac{dz}{L}
                   \int_0^{2\pi}\frac{\cos^2\varphi}{2\pi}\,d\varphi\int_0^{+\infty}r^2
                   u'(r)u(r)\,dr\\
                        &=&\alpha\int_0^{+\infty}r^2 u'(r)u(r)\,dr
\end{eqnarray*}
In a previous paper~\cite{prstab} we adopted the uncertainty product
$\Delta x\cdot\Delta p_x$ as a measure of the r.m.s.\ emittance. As
an example let us suppose again that our wave function has the form
$u_0(r)$ for the harmonic oscillator without space charge given
in\refeq{unpertground}. We then have
\begin{equation}
  \Delta x\cdot\Delta p_x=-C=\alpha\,.\label{corrcoeff}
\end{equation}
This allows two remarks: first, $\alpha$ plays also the role of a
measure of the emittance and hence -- as suggested in a previous
paper~\cite{pre} -- its value must be linked to the number of
particles in the beam; second, the position--momentum correlation
coefficient of a Gaussian beam is
\begin{equation*}
  \frac{C}{\Delta x\cdot\Delta p_x}=-1
\end{equation*}
as it was predictable, since in SM the relation between position and
momentum for the wave function\refeq{unpertground} is linear and
negative.

In other models the transverse r.m.s.\ emittance is calculated by
means of the quantity $\sqrt{\Delta x^2\,\Delta p_x^2-C^2}$. In the
KV distribution, since momentum and position are uncorrelated and
$\langle x\rangle=\langle p_x\rangle=0$, this estimate becomes
$\sqrt{\langle x^2\rangle\langle p_x^2\rangle-\langle x
p_x\rangle^2}$. In the SM model, on the contrary, this is not a good
choice: in fact we have shown, at least in our simple example, that
$x$ and $p_x$ are far to be uncorrelated, and that as a consequence
of\refeq{corrcoeff} $\sqrt{\Delta x^2\,\Delta p_x^2-C^2}$ becomes
exactly zero. Apparently it is not realistic to take this value as a
good estimate of the emittance. On the other hand, for the same
gaussian example, the value of the uncertainty product $\Delta
x\,\Delta p_x$ is just $\alpha$ which we assume to be a good
candidate for the value of the emittance. On the other hand it is
easy to calculate the same uncertainty product for a Student
distribution $\Sigma_2(\nu,(\nu-2)\sigma^2)$ with dimensionless
radial distribution\refeq{dlradialstudent} and variance $\sigma^2$:
in fact a straight application of\refeq{deltax} and\refeq{deltap}
brings to the following result
\begin{equation}\label{studentemitt}
  \Delta x\cdot\Delta p_x=\alpha\,\sqrt{\frac{\nu(\nu+2)}{(\nu-2)(\nu+4)}}
\end{equation}
Of course, as it is already clear, this value converges to the
Gaussian case for large $\nu$, while becomes larger and larger for
small $\nu$ values when the shape of the distribution moves away
from the Gaussian case.

\subsection{Weighing the tails}

We can finally compare the length of the tails of Gauss and Student
distribution in order do assess the possible halo formation in the
second case. Let us consider the probability
\begin{equation}\label{probbeyond}
    P(c)=\int_{c\sigma}^{+\infty}ru^2(r)\,\mathrm{d}r
\end{equation}
of being beyond a distance $c\sigma$ ($\sigma^2$ being the variance)
away from the beam axis, and calculate this quantity in our two
cases. From the Gaussian distribution we have
from\refeq{unpertground} that
\begin{equation}\label{probbeyondgauss}
    P(c)=e^{-c^2/2},
\end{equation}
while in the Student case from\refeq{3DbeamstudentVar} we get
\begin{equation}\label{probbeyondstud}
    P(c)=\left(1+\frac{c^2}{\nu-2}\right)^{-\nu/2}.
\end{equation}
Now for $c=10$ the Gaussian value is about $1.9\times 10^{-22}$,
while with $\nu=10$ the Student value is about $2.2\times 10^{-6}$,
and with $\nu=22$ the value is $2.8\times 10^{-9}$. This means that
for $\mathcal{N}=10^{11}$ particle per meter of beam, we find
practically no particle beyond $10\sigma$ in the Gaussian case, but
about $10^3$ particle per meter for a $\nu=22$ Student distribution,
and as much as $10^5$ for a $\nu=10$ value. It is worthwhile to
remember at this point that we got about the same number of
particles gone astray in our self--consistent numerical solutions
for a dimensionless perveance of about $\xi=20$ in one of our
previous paper~\cite{prstab}.

\section{L\'evy--Student processes}\label{studentprocess}

In our context the Student laws $\Sigma(\nu,a^2)$ are important not
only because they promise to better describe the halo by means of
their longer tails with respect to usual Gaussian distributions; in
fact they constitute an important family of L\'evy \emph{infinitely
divisible} (i.d.)\ laws. At present there is a lot of interest about
non--Gaussian L\'evy laws in several fields of research (see for
example~\cite{paul,mantegna} and references quoted therein), but
this interest is mostly confined to the \emph{stable} laws which are
in fact an important sub--family of the i.d.\ laws. The fundamental
character of the i.d.\ laws can be better understood from two
different, but strictly correlated standpoints: on the one hand the
i.d.\ laws constitute the more general form of possible limit laws
for the generalized Central Limit Theorem; on the other they
constitute the class of all the laws of the increments for every
stationary, stochastically continuous, independent increments
process (L\'evy process). These important results (which are briefly
discussed in Appendix~\ref{idstlaws} and Appendix~\ref{CLT}) have
been achieved by P.\ L\'evy, A.Ya.\ Khintchin, A. Kolmogorov and
other mathematicians from the mid 30's to the mid 40's of the XXth
century, but their relevance for the applications has been
recognized only in more recent years. One of the characteristics of
a non--Gaussian L\'evy process is to have trajectories with moving
discontinuities (think to the trajectories of a typical Poisson
process contrasted with those of a Gaussian Wiener process), and we
propose here to describe the trajectories of the particle beam by
means of a L\'evy--Student process whose discontinuities can
possibly account for the relatively rare escape of particles from
the beam core. For the sake of simplicity we will limit ourselves in
the following to the case of 1--DIM systems representing one single
transverse coordinate of our particle beam.

\subsection{The Student i.d.\ laws}\label{studentid}

The ch.f.'s of the laws $\Sigma(\nu,a^2)$, namely the Fourier
transform of the densities\refeq{1Dstudentpdf}, are
\begin{figure}
\begin{center}
\includegraphics*[width=8.0cm]{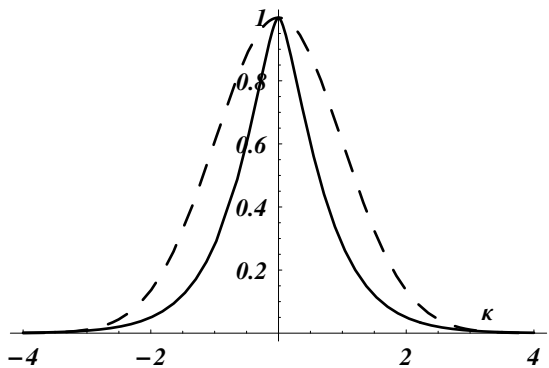}
\end{center}
\caption{Typical ch.f. of a Student law $\Sigma(2,2)$ (solid line)
compared with the ch.f. of a standard Gauss law $\mathcal{N}(0,1)$
(dashed line).}\label{chf}
\end{figure}
\begin{equation}\label{1Dstudentchf}
    \varphi(\kappa)=2\,\frac{|a\kappa|^{\frac{\nu}{2}}\mathrm{K}_{\frac{\nu}{2}}(|a\kappa|)}
    {2^{\frac{\nu}{2}}\Gamma\left(\frac{\nu}{2}\right)}
\end{equation}
where $\mathrm{K}_\alpha(z)$ is a modified Bessel function. The
typical form of these ch.f's (contrasted with the Gauss ch.f.)\ is
shown in FIG.\myref{chf}. Remark that, since $x$ is a length, the
ch.f.\ variable $\kappa$ has the dimensions of wave number (inverse
of a length). These laws are i.d.\ but in general are not stable,
the unique stable laws among them being the Cauchy laws
$\mathcal{C}(a)=\Sigma(1,a^2)$. For $\nu>2$ the Student
distributions belong to the domain of attraction of the Gauss law
since they have a finite variance. For $\nu\leq2$ the variance
diverges, but it is possible to prove that this notwithstanding they
still belong to the Gaussian domain also for $\nu=2$: albeit this
derives from a well known general result~\cite{bouchaud} a simple
proof for our particular case will be given in a subsequent paper.
On the other hand for $\nu<2$ the Student laws are attracted by
non--Gaussian stable laws characterized by the same value of the
parameter $\nu$.

The fact that the Student laws are i.d.\ -- which in itself is not
at all a trivial result proved in steps only in the 70's and
80's~\cite{bondesson} -- shows two kinds of advantages with respect
to more common stable laws:
\begin{itemize}
    \item no stable, non--Gaussian law can have a
    finite variance, while all Student laws with $\nu>2$ do have a
    finite variance; this is important since it is not
    realistic to suppose that empirical distributions (in particular for the particle beams) have infinite
    variances, but that notwithstanding we will not be obliged to
    resort to handmade modification
    (for instance \emph{truncated} L\'evy distributions) as in the case of stable distributions~\cite{paul};
    \item the asymptotic behavior of stable, non--Gaussian laws is
    proportional to $|x|^{-\alpha-1}$ with $\alpha<2$~\cite{mantegna}, while the
    asymptotic behavior of the Student laws is $|x|^{-\nu-1}$ with
    $\nu>0$; this allows the Student laws -- but not the stable laws -- to continuously go throughout all the gamut
    of decay speeds to approximate in a fine tuning the Gaussian behavior as well
    as we want.
\end{itemize}
The principal drawback for not being stable is in the subsequent
definition of the L\'evy--Student process. In fact the ch.f.\ of the
process $\varphi(\kappa,t)$ coincides with\refeq{1Dstudentchf} only
for $t-s=T$, while for $t-s\neq T$ it is no more the ch.f.\ of a
$\Sigma(\nu,a^2)$ law. Hence we explicitly know the form of the
increment law only at the time scale $T$: we know the ch.f.\ --
namely everything we theoretically need -- at every time, but we do
not have the explicit inverse Fourier transform, and we also know
that the laws are no more in the family $\Sigma(\nu,a^2)$. This
problem is tempered by the remark that the situation is not better
for general stable laws: even in this case, in fact, we do not know
the explicit forms of the increment laws not even for one time scale
(they are known only in a few precious instances). The unique
advantage in the stable case being the fact that all along the time
evolution the increment laws remain of the same type, which is not
the case for i.d.\ non stable laws.

\subsection{The L\'evy--Student process}

\begin{figure}
\begin{center}
\includegraphics*[width=8.0cm]{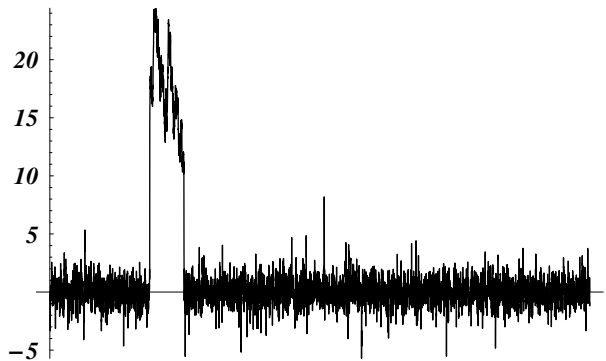}
\end{center}
\caption{Occasional trajectory of a stationary, Student process
$Y(t)$ (see EDS~\refeq{studentEDS}, $\nu=4$) with a temporary
excursion out of the core.}\label{student2}
\end{figure}

\begin{figure}
\begin{center}
\includegraphics*[width=8.0cm]{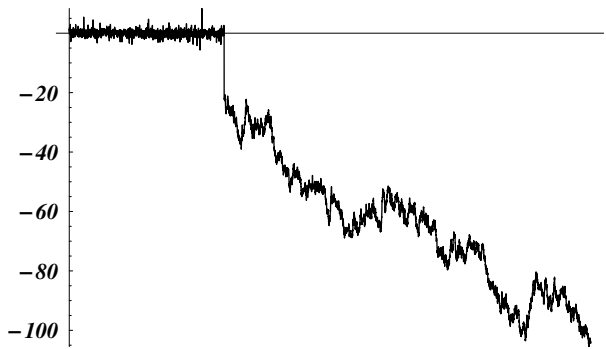}
\end{center}
\caption{Rare, but possible trajectory of a stationary, Student
process $Y(t)$ (see EDS~\refeq{studentEDS}, $\nu=4$): here the
particle definitely drifts away from the core.}\label{student3}
\end{figure}

A L\'evy process defined by the ch.f.\refeq{1Dstudentchf} will be
called in the following a L\'evy--Student process. Taking into
account\refeq{levytpdf} and\refeq{1Dstudentchf} we can now state
that the transition pdf of a L\'evy--Student process is
\begin{eqnarray}\label{studenttpdf}
    \lefteqn{p(x,t|y,s)=} \nonumber\\
    & & \frac{1}{2\pi}\int_{-\infty}^{+\infty}e^{i\kappa(x-y)}
                   \left[2\frac{|a\kappa|^{\frac{\nu}{2}}\mathrm{K}_{\frac{\nu}{2}}(|a\kappa|)}
    {2^{\frac{\nu}{2}}\Gamma\left(\frac{\nu}{2}\right)}\right]^{\frac{t-s}{T}}\!\!\mathrm{d}\kappa
\end{eqnarray}
where the improper integral is always convergent since the
asymptotic behavior of the ch.f.\ is
\begin{equation*}
    \varphi(\kappa)=\sqrt{2\pi}\,\,\frac{|a\kappa|^{\frac{\nu-1}{2}}e^{-|a\kappa|}\left[1+\mathrm{O}(|\kappa|^{-1})\right]}
                                               {2^{\frac{\nu}{2}}\Gamma\left(\frac{\nu}{2}\right)}\,,\quad
    |\kappa|\rightarrow+\infty
\end{equation*}
In principle\refeq{studenttpdf} should be enough to calculate
everything of our process, but in practice this is an integral that
must be treated numerically, but for a few particular cases that
will be discussed in a subsequent paper. On the other hand even to
produce simulation for the trajectories of our process we should
have some simple expression for the transition pdf. At least for
this last task, however, we can exploit the fact that when $t-s=T$
the expression\refeq{studenttpdf} can be exactly calculated and
coincides with the pdf\refeq{1Dstudentpdf} of a Student
$\Sigma(\nu,a^2)$ (remark that even this is not possible for the
typical non--Gaussian stable process). This means that we can
produce sample trajectories by taking $T$ as the fundamental step of
our numerical simulation. In other words we will simulate the sample
paths of a process whose increments are exactly Student distributed
when observed at the (otherwise arbitrary) time scale $T$. To give a
look to these trajectories we produced a simplified model which
simulates the solutions of the following two SDE's
\begin{figure}
\begin{center}
\includegraphics*[width=8.0cm]{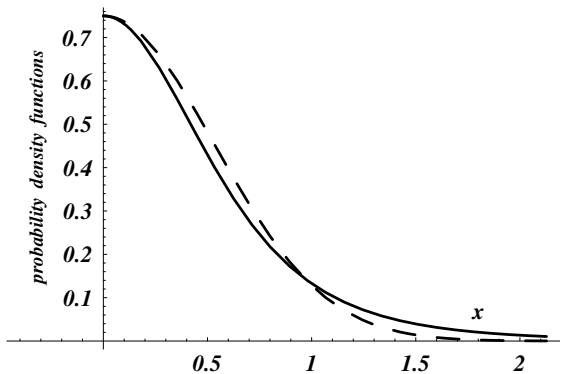}
\end{center}
\caption{The pdf's of the increments for the Gaussian processes
(SDE\refeq{wienerEDS}, dashed line; $\sigma\simeq0.53$) and for the
L\'evy--Student process with law $\Sigma(4,1)$
(SDE\refeq{studentEDS}, solid line; $\sigma\simeq0.71$). The
parameters are chosen so that the two pdf's have the same modal
values and similar shapes.}\label{GaussStud2}
\end{figure}
\begin{eqnarray}
  \mathrm{d}X(t) &=& v(X(t),t)\,\mathrm{d}t + \mathrm{d}W(t)\label{wienerEDS}\\
  \mathrm{d}Y(t) &=& v(Y(t),t)\,\mathrm{d}t +
  \mathrm{d}S(t)\label{studentEDS}
\end{eqnarray}
where $W(t)$ is a Wiener process, while $S(t)$ is a L\'evy-Student
process. We also fixed the velocity field $v(x,t)$ in a suitable
way: it will not depend on time $t$, and its value is (for given
$b>0$ and $q>0$)
\begin{figure}
\begin{center}
\includegraphics*[width=8.0cm]{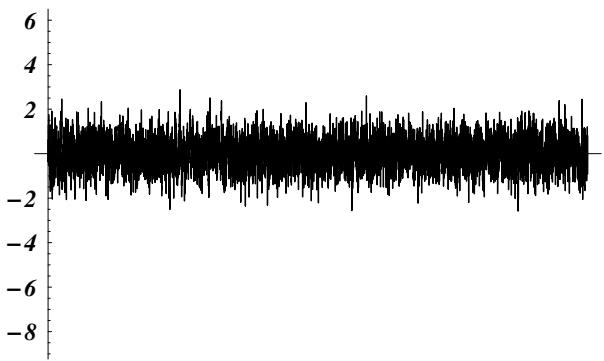}
\end{center}
\caption{Typical trajectory of a stationary, Gaussian
(Ornstein--Uhlenbeck) process $X(t)$ (see SDE~\refeq{wienerEDS}). To
compare it with the Student trajectory, the vertical scale has been
set equal to that of FIG.\myref{student1}}\label{gauss1}
\end{figure}
\begin{figure}
\begin{center}
\includegraphics*[width=8.0cm]{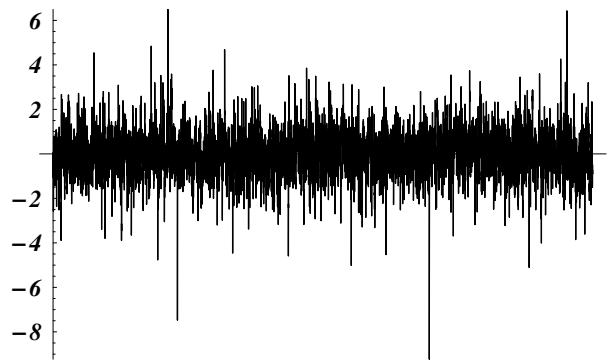}
\end{center}
\caption{Typical trajectory of a stationary, Student process $Y(t)$
(see EDS~\refeq{studentEDS}, $\nu=4$ and $a=1$).}\label{student1}
\end{figure}
\begin{equation*}
    v(x)=-bx\,H(q-|x|)
\end{equation*}
where $H$ is the Heaviside function. This flux will attract the
trajectory toward the origin when $|x|\leq q$, and will allow the
movement to be completely free for $|x|>q$. The forms of the typical
pdf's used in our simulations are shown in FIG.\myref{GaussStud2}.
In a simplified model for a collimated beam this will then produce a
stationary, Ornstein--Uhlenbeck process for the SDE\refeq{wienerEDS}
if the intensity of the Gaussian noise is not too large. The process
solution of the SDE\refeq{studentEDS} will instead have different
characteristics. Let us suppose to fix the ideas that the two
parameters defining the velocity field are $b=0.35$ and $q=10$. The
FIG.\myref{gauss1} displays a typical trajectory of a $10^4$ steps
solution $X(t)$ of\refeq{wienerEDS} when the variance of the
Gaussian distributed increments is $\sigma^2=0.28$. In our
simplified 1--DIM model of the transverse dynamics of a particle
beam this means that the trajectories always stay inside the beam
core. Let us then take as law for the increments
of\refeq{studentEDS} a Student distribution $\Sigma(4,1)$: its pdf
looks not very different from that of the previous Gaussian
distribution, as the FIG.\myref{GaussStud2} clearly show. That
notwithstanding the process $Y(t)$ differs in several respects from
$X(t)$. Indeed not only the typical trajectory displayed in
FIG.\myref{student1} shows a wider dispersion of its values and a
few larger spikes. The principal difference is rather in the fact
that while the trajectories of $X(t)$ show a remarkable stability in
their statistical behavior, the paths of $Y(t)$ have the propensity
to make occasional excursions far away from the beam core (see
FIG.\myref{student2}), and seldom they also definitely drift away
from the core (see FIG.\myref{student3}). This depends of course on
the mentioned properties of the trajectories of a non--Gaussian
L\'evy process, and in particular on the fact that they are only
stochastically, and not pathwise continuous, namely that they
contain occasional jumps. The frequency and the size of these jumps
can also be fine tuned by suitably choosing the values of the
parameters of the law $\Sigma(\nu,a^2)$ of the increments. It is
this feature of a L\'evy--Student process that suggests to adopt
this model to describe the rare escape of particles away from the
beam core.

\section{Conclusions}\label{conclusions}

In the previous sections we have introduced the Student laws in our
SM model for the particle beam dynamics first of all in order to
make use of their features depending on their enhanced variance. In
particular we have shown that the longer tails with respect to the
similar Gaussian distributions can help to account for the finding
of a larger than expected number of particles removed far away from
the beam core.

It should be remarked, however, that all along the
Section\myref{selfconspot} our processes were Gaussian processes
since the underlying SDE\refeq{ito} is still powered by a Brownian
noise. This is true even in the Section\myref{studtransvdist} where
we first introduced the Student laws\refeq{1Dstudentpdf} as
stationary distributions of the process. It is only in the
Section\myref{studentprocess} that we introduced a new kind of SDE
with a L\'evy--Student noise. The more relevant feature of these
processes is the fact that their trajectories make jumps: indeed
this can become a model for the halo formation in the beams. From a
physical point of view these jumps can be produced by occasional
hard collisions among the beam particles, the probability of these
collisions growing with the intensity of the beam. In some sense it
is not only the variance of the transverse distribution of the beam
which principally rules the emergence of a halo: in the simulations
produced here the variances of the Gaussian and of the Student
processes were roughly the same. Rather it is the qualitative
character of the process which accounts for the rare escape of the
particles from the beam core. For a process produced by a Gaussian
noise (a process pathwise continuous: almost every trajectory is
everywhere continuous) there is no chance to observe trajectories
going out of a well collimated beam. On the contrary, for a process
produced by a L\'evy--Student noise (a process only stochastically
continuous: trajectories can have jumps) occasionally the jump is
large enough to put the particle out of the stream. Of course the
frequency and the size of these jumps depend on the parameters $\nu$
and $a$ of the process: the jumps tend to be smaller and less
frequent when the $\Sigma(\nu,a^2)$ distributions approximate a
Gaussian law. In our opinion it would be very interesting to explore
the possibility that the processes underlying the intense beam
dynamics be ruled by some sort of L\'evy--Student noise rather than
by the usual Gaussian noise. It is then important to point out that
a few numerical evidences~\cite{vivoli} begin to emerge which
confirm this conjecture.

These remarks point to several research directions. First of all it
is important to better study the L\'evy--Student process in itself:
for example a knowledge of the L\'evy--Khintchin functions of the
Student laws would be relevant to the fine tuning of the frequency
and the size of the trajectory jumps. On the other hand even the
differential form of its Chapman--Kolmogorov
equation~\cite{gardiner} would be instrumental to discuss the time
evolution of the process. Then it must be remarked that at present
we have just defined the L\'evy--Student process, but we added no
dynamics: it is as if we have the Wiener process, but no Stochastic
Mechanics or any other dynamical model added to this kinematics. In
other words we need to build a new generalized SM for the
L\'evy--Student processes. Finally it would be important at this
point to have empirical or numerical data able to corroborate the
hypothesis that the increments of the transverse variables of a beam
are in fact distributed according to a Student law, rather than
according to the usual Gaussian law.

\begin{acknowledgments}
We would like to thank Dr. C. Benedetti,  Prof. F. Mainardi,  Prof.
G. Turchetti and Dr. A. Vivoli for useful comments and suggestions.
\end{acknowledgments}

\appendix

\section{Infinitely divisible and stable laws}\label{idstlaws}

The relevant mathematical concepts used in this paper are better
discussed in the framework of the theory of the addition of
independent random variables (r.v.): for more details
see~\cite{loeve,feller,gnedenko}. In the following we will describe
the law $\mathcal{L}$ of a r.v.\ $X$ by giving her characteristic
function (ch.f.)
\begin{equation*}
    \varphi(\kappa)=\mathbf{E}(e^{i\kappa X})
\end{equation*}
where $\mathbf{E}(\cdot)$ is the expectation under the law
$\mathcal{L}$. When $\mathcal{L}$ has a pdf $f(x)$, then
$\varphi(\kappa)$ is just its Fourier transform. It is well known
that the law $\mathcal{L}$ of the sum of $n$ independent r.v.'s with
laws $\mathcal{L}_1,\ldots,\mathcal{L}_n$ has a ch.f.\ which is the
product of the ch.f.'s of the component laws:
\begin{equation}\label{addrv}
    \varphi(\kappa)=\varphi_1(\kappa)\cdot\ldots\cdot\varphi_n(\kappa)
\end{equation}
On the other hand we say that a law $\mathcal{L}$ is
\emph{decomposed} in the laws $\mathcal{L}_1,\ldots,\mathcal{L}_n$
when its ch.f.\ can be written as a product\refeq{addrv} of the
ch.f.'s of its components. This already allows us to introduce two
fundamental concepts: a law $\mathcal{L}$ with ch.f. $\varphi$ is
said to be i.d.\ when for every $n$ there is a law $\mathcal{L}_n$
with ch.f.\ $\varphi_n$ such that $\varphi=\varphi_n^n$. In other
words this means that for every $n$ a r.v.\ $X$ with law
$\mathcal{L}$ can always be decomposed in the sum of $n$ independent
r.v.'s all with the same law $\mathcal{L}_n$ (identically
distributed). Remark, however, that in general the laws
$\mathcal{L}_n$ are not of the same type as $\mathcal{L}$. Let us
remember here that we say that two laws are of the same \emph{type}
when we get one from the other by means of a centering and a
rescaling; in other words, if $\varphi(\kappa)$ in a ch.f., then all
the ch.f.'s of the same type have the form
$e^{ia\kappa}\varphi(b\kappa)$ for every $a$ and $b>0$. For instance
all the Gaussian laws $\mathcal{N}(\mu,\sigma^2)$ belong to the same
(Gaussian) type; on the contrary the Poisson laws
$\mathcal{P}(\lambda)$ with different values of $\lambda$ do not
belong to the same type. Now, a law $\mathcal{L}$ is said to be
stable when it is i.d.\ and the component laws are of the same type
as $\mathcal{L}$. More precisely a ch.f.\ $\varphi(\kappa)$ is
stable when for every $b, b'>0$ there exist $a$ and $c$ such that
\begin{equation*}
    \varphi(c\kappa)=e^{ia\kappa}\varphi(b\kappa)\varphi(b'\kappa)\,.
\end{equation*}
As an example: the Gaussian and the Cauchy laws are stable; the
Poisson laws are instead only i.d. The families of i.d.\ and stable
laws are completely characterized: in fact the celebrated
L\'evy--Khintchin formula gives the more general form for the
ch.f.'s of these two classes; however, while in the case of the
stable laws these ch.f.'s (albeit not in general the laws
themselves) are explicitly known in terms of elementary functions,
for the i.d.\ laws the ch.f.'s are given through an integral
containing a function $L(x)$ (\emph{L\'evy function}) associated to
every particular law. But for a few classical cases the L\'evy
functions of the i.d.\ laws are not known.

\section{Central Limit Theorem and L\'evy processes}\label{CLT}

Let us consider the sequence of r.v.'s $X_{n,k}$ with
$n\in\mathbb{N}$ and $k=1,\ldots,n$ with $X_{n,1},\ldots,X_{n,n}$
independent for every $n$. The modern formulation of the Central
Limit Problem asks to find the more general laws which are limits of
the laws of the \emph{consecutive sums}
\begin{equation}\label{consecsum}
    S_n=\sum_{k=1}^nX_{n,k}
\end{equation}
Remark that these sums generalize the usual partial sums of the
classical Central Limit Theorem in that: when we go from $S_n$ to,
say, $S_{n+1}$, the first $n$ terms do not in general remain the
same: for example $X_{n.1}$ does not coincide with $X_{n+1.1}$.
Under very general technical conditions the Central Limit Theorem
now states that the family of all the limit laws of the consecutive
sums\refeq{consecsum} coincides with the family of i.d.\ laws. The
stable laws come into play only when we specialize the form of our
consecutive sums: when we have
\begin{equation*}
    X_{n,k}=\frac{X_k}{a_n}-\frac{b_n}{n}
\end{equation*}
where $a_n$ and $b_n$ are sequences of numbers, and $X_k$ are
independent r.v.'s, the consecutive sums take the form of the usual
\emph{normed sums} (centered and rescaled sums of independent
r.v.'s)
\begin{equation}\label{normsum}
    S_n=\frac{S^*_n}{a_n}-b_n\,,\qquad\quad S^*_n=\sum_{k=1}^nX_k\,.
\end{equation}
Then, if the $X_k$ are also identically distributed, the family of
the limit laws of the normed sums\refeq{normsum} coincides with the
family of the stable laws. The classical (Gaussian) Central Limit
Theorem is an example of convergence toward a stable law; on the
other hand the Poisson Theorem (convergence of Binomial laws toward
Poisson laws) is an example of convergence toward an i.d.\ law.
Every stable law has its own domain of attraction, namely the set of
laws attracted by it in the sense of the convergence of normed
sums\refeq{normsum} of independent r.v.'s all distributed as the
attracted law. It can be proved that all the laws with finite
variance are in the domain of attraction of the Gauss law, and that
a law can be attracted by a non--Gaussian stable law only if it has
infinite variance.

The general formulation of the Central Limit Theorem is strictly
connected to the definition of the processes with independent
increments (\emph{decomposable processes}). It is apparent in fact
that if the increments $\Delta X(t)=X(t+\Delta t)-X(t)$ for non
superposed intervals are independent, the previous forms of the
Central Limit Theorem imply that the laws of the increments must be
i.d.\ laws. Moreover, since the decomposable process are also Markov
processes, the laws of the increments are also all that is needed to
completely define them. If a decomposable processes $X(t)$ is
\emph{stationary} (namely the law of $X(t+s)-X(s)$ does not depend
on $s$) and \emph{stochastically continuous} (namely for every $t$
we have $X(t+\Delta t)-X(t)\rightarrow0$ in probability when $\Delta
t\rightarrow0$) we will call it a \emph{L\'evy process}. Remark that
a Poisson process is a L\'evy process since, despite its
discontinuities, it is stochastically continuous. In fact these
discontinuities do not impair the stochastic continuity of the
process because they are \emph{moving} (as opposed to \emph{fixed})
discontinuities. On the other hand it is possible to prove that only
the Gaussian L\'evy processes (for example the Wiener, or the
Ornstein--Uhlenbeck processes) are \emph{pathwise continuous},
namely: almost every sample path is everywhere continuous (there are
not even moving discontinuities). Now, if $\varphi(\kappa)$ is the
ch.f.\ of an i.d.\ law and $T$ is a suitable time constant, it is
possible to prove that $[\varphi(\kappa)]^{\Delta t/T}$ is the
ch.f.\ of the increments $\Delta X(t)$ of a L\'evy process. Hence,
if the process has a pdf, the stationary transition pdf is
\begin{equation}\label{levytpdf}
    p(x,t|y,s)=\frac{1}{2\pi}\,\mathrm{PV}\int_{-\infty}^{+\infty}e^{i\kappa(x-y)}
                   [\varphi(\kappa)]^{\frac{t-s}{T}}\mathrm{d}\kappa
\end{equation}
so that, at least in principle, we know all that is needed to define
the process.

The sample paths of a L\'evy process are also well characterized: it
is possible in fact to prove that almost all trajectories are
bounded and are continuous with the exception of a countable set of
moving jumps (first kind discontinuities). Then, let us suppose that
$L_t(x)$ is the L\'evy--Khintchin function of the i.d.\ law of the
increment $X(s+t)-X(s)$: if $\nu_t(x)$ is the random number of the
jumps in $[s,s+t)$ of height in absolute value larger than $x>0$, it
is possible to prove that
\begin{equation*}
    |L_t(x)|=\mathbf{E}(\nu_t(x))
\end{equation*}
so that the L\'evy--Khintchin function of an i.d.\ law plays also
the role of a measure of the frequency and height of the trajectory
jumps.

\vfill\eject

\end{document}